\documentclass[12pt]{article}
\usepackage{amsmath}
\usepackage{cite}
\def\MSbar{\relax\ifmmode\overline{\rm MS}\else{$\overline{\rm MS}${ }}\fi}
\def\smallMSbar{\mbox{\tiny ${\rm  MS}$}}
\def\almsbar{\alpha_{\smallMSbar}}
\def\smallPh{\mbox{\tiny ${\rm  Ph}$}}
\def\alphys{\alpha_{\smallPh}}
\def\gone{\gamma_{\mbox{\tiny 1}}}
\def\dgone{{\dot \gamma}_{\mbox{\tiny 1}}}
\def\ddgone{{\ddot \gamma}_{\mbox{\tiny 1}}}

\def\eV{{\rm e\kern-0.12em V}}            
  
\def\half{{\textstyle {\frac12}}}

\newcommand{\al}{\alpha}

\def \as{\relax\ifmmode\alpha_s\else{$\alpha_s${ }}\fi}

\def\np#1#2#3{{Nucl.\ Phys.\ }{\bf B#1} (#3) #2}

\def\hG{\partial_{t}}

\def\Li2{\mathop{\rm Li_2}}
\def\cO#1{{\cal{O}}\!\left(#1\right)}
\def\cP{{\cal{P}}}

\title{\textbf{Revisiting parton evolution and the large-$x$ limit}}
\author{Yu.~L.\ Dokshitzer$^1$\footnote{On leave of absence: St.\ Petersburg Nuclear Physics Institute, 188350, Gatchina, Russia},
  G.~Marchesini$^{2,1}$ and 
  G.~P.~Salam$^1$ \\[3pt] 
  \normalsize
$^1$LPTHE, Universities of Paris-VI and VII and CNRS, Paris, France\\
\normalsize
$^2$University of Milano--Bicocca and INFN Sezione di Milano, Milan, Italy
}
\date{}

\begin{document}

\maketitle

\vspace{-7.3cm}
\begin{flushright}
  Bicocca--FT--05--31\\
  LPTHE--05--31\\
  hep-ph/0511302~v2
\end{flushright}
\vspace{4.5cm}

\abstract{This remark is part of an ongoing project to simplify the
  structure of the multi-loop anomalous dimensions for
  parton distributions and fragmentation functions. It
  answers the call for a ``structural explanation'' of a ``very
  suggestive'' relation found by Moch, Vermaseren and Vogt in the
  context of the $x\!\to\! 1$ behaviour of three-loop DIS anomalous
  dimensions. It also highlights further structure that remains to be
  fully explained.}

\section{Introduction} 

This letter stems from a project to better understand the structure of
multi-loop anomalous dimensions both for parton distributions and
fragmentation functions~\cite{DMSup}. These distributions, which we
shall generically denote as $D$ ($D_N(Q^2)$ in moment space, $D(x,Q^2)$ in
$x$-space) satisfy a renormalisation group equation
\begin{equation}
  \label{eq:anom-dim}
  \frac{d D_N(Q^2)}{ d \ln Q^2} \equiv \hG D =
  \gamma(N, \as(Q^2)) D_N(Q^2)\,,
\end{equation}
where $\hG$ is a compact notation for the derivative with respect to
$t=\ln Q^2$ and $\gamma(N,\as)$ are elements of an anomalous dimension
matrix.  The latter have been calculated in terms of an expansion in
the coupling $\as$ up to three (two) loops in the space-like
(time-like) cases \cite{MVV,VMV,CFP,KKST}. They become increasingly
cumbersome beyond leading order.

Conventionally one defines parton splitting functions, $P(x)$, as the
inverse Mellin transform of the corresponding anomalous dimensions,
giving evolution equation in $x$ space in terms of a direct convolution 
\begin{equation}
  \label{eq:conventional-split}
  \hG D(x,Q^2) = \int_0^1 \frac{dz}{z} P(z,\as(Q^2)) \,
              D\left(\frac{x}{z},Q^2\right)\,,
\end{equation}
where $D(x,Q^2)$ has the physical support $x\le 1$. We have reason to
suspect that there might exist a reformulation of the evolution
equations (\ref{eq:conventional-split}) in which, by generalising the
structure on the right-hand-side, one is able to simplify the
splitting functions. This is equivalent to stating that the
higher-loop structure of the anomalous dimensions in
(\ref{eq:anom-dim}) can in part be understood as inherited from
non-linear combinations of lower loops.

In this picture, the new splitting functions would not only be more
compact, but they would also exhibit some important physical
properties: beyond first loop they should vanish at large $x$, and
they should be identical for space-like and time-like evolution, thus
restoring Gribov-Lipatov reciprocity~\cite{GL} (broken beyond first
loop in the standard formulation, see \cite{CFP}).

A possible reformulation of (\ref{eq:conventional-split})
is\footnote{Such a reformulation of the notion of parton splitting
  functions originally proposed in~\cite{Dunpub} has been carried out
  in detail in the context of heavy-quark fragmentation functions,
  where it was found to greatly improve the perturbative
  series \cite{DKTheavy}.} %
\begin{equation}
  \label{eq:new-split}
  \hG D(x,Q^2) = \int_0^1 \frac{dz}{z} \cP(z,\as(z^{-1} Q^2)) \,
              D\left(\frac{x}{z},z^\sigma Q^2\right)\,,
\end{equation}
where $\sigma = -1$ ($+1$) for the space-like (time-like) case. The
choice of $z^{\sigma}Q^2$ as the logarithmic ordering parameter
(parton evolution `time') corresponds to ordering parton splittings in
the \emph{fluctuation lifetimes} of successive virtual parton states.
Assuming the new $\cP(x)$ splitting functions to be identical in the
space-like and time-like cases one obtains that the `traditional'
two-loop splitting functions in the two cases, $P^{(2),S}(x)$ and
$P^{(2),T}(x)$
differ by
\begin{equation}\label{GLviol} 
 \half( P^{(2),T}_{ns} - P^{(2),S}_{ns} ) = \int_0^1
 dz \int_0^1 dy \,\delta(x-yz) \>P_{qq}^{(1)}(z)\ln z \cdot
 \left\{P^{(1)}_{qq}(y)\right\}_{+}\, .  
\end{equation} 
This is precisely the relation that was noted by Curci, Furmanski and
Petronzio in \cite{CFP} for non-singlet quark evolution. In the
singlet case (both for quarks and gluons) there are also interesting
patterns, see appendix. Furthermore, if one writes $\cP(z,\as)$ as a
series in the physical coupling, $\alphys = \alpha_{\smallMSbar} +
\frac{K}{4\pi} \almsbar^2 + \cdots$ (with $K =
(\frac{67}{9}-\frac{\pi^2}{3})C_A - \frac{10}{9}n_f$), the two-loop part of $\cP(x,\as)$
vanishes as $(1-x)$ for $x\to 1$.
This corresponds to the wisdom of Low, Barnett and Kroll \cite{LBK}
according to which the classical nature of soft radiation reveals
itself at the level of the $1/(1-x)$ and constant terms. This
classical nature of soft radiation allows one to absorb all soft
singularities into the first loop and to look upon higher-loop
splitting functions as due to true multi-parton (quantum)
fluctuations.

Finally, one notes also the $z$-dependence of the argument of the
coupling in (\ref{eq:new-split}). Its trace is visible in the explicit
structure of all the diagonal two-loop anomalous dimensions. Moreover,
this argument naturally emerges when using dispersive reasoning to
carry out a careful treatment of the appearance of the running
coupling in inclusive processes \cite{DS}.

We are still far from a good understanding of how to simplify the
structure of multi-loop anomalous dimensions, notably because of
complications that arise from off-diagonal transitions (see appendix).
Nevertheless the belief that the $x\to 1$ limit is under control has
led us to investigate the implications of eq.~(\ref{eq:new-split}) for
the large-$N$ structure of three loop anomalous dimensions.

As we shall see this will provide insight concerning a ``{\em very
  suggestive}'' relation noted by Moch, Vermaseren and Vogt (MVV)
which, in their words, ``{\em seems to call for a structural
  explanation}'': in both the non-singlet quark \cite{MVV} and
diagonal singlet quark-quark and gluon-gluon three-loop splitting
functions \cite{VMV}, they observed that the third-loop coefficients,
$C_3^a$, in the large-$N$ expansion of the $n$-loop anomalous
dimensions,\footnote{We define $\gamma_{n}(N) = \int_0^1 dz z^{N-1}
  P^{(n)}(z)$; this has the opposite sign to the convention of MVV; we
  have also changed the sign in front $C^a$ as compared to eq.~(3.10)
  of ref.~\cite{MVV}, which contains a misprint.}
\begin{multline}
  \label{MVVlargeN} \gamma_{aa} (N) = - A^a(\ln N +\gamma_e) +
  B^a - C^a \,N^{-1} \ln N \>+\>\cO{N^{-1}} \,,
  \\
  \gamma_{aa} \equiv \sum \gamma_{n,aa}
  \left(\frac{\almsbar}{4\pi}\right)^n\,,
  \qquad
  A^a \equiv \sum A_{n}^a \left(\frac{\almsbar}{4\pi}\right)^n\,,
  \quad\mbox{etc.}
\end{multline} 
are simply related with $A_2^a$ and $A_1^a$, i.e.\ $C_3^a = 2 A_1^a
A_2^a$, where $a = q,g$. This supplements the two-loop relation, $C_2^a =
(A_1^a)^2$ \cite{CFP}.

\section{MVV relation} 

For the purpose of studying the $x\to 1$ limit we initially
approximate $\cP(x,\as)$ by the product of the physical coupling,
$\alphys$, and the 1-loop splitting function, $\cP(x,\as) \simeq
(\alphys/4\pi) P^{(1)}(x)$ (for compactness we will write
$(\alphys/4\pi) \equiv \al$).  To deal with the correlated $z$ and
$Q^2$ dependences in the right-hand side of (\ref{eq:new-split}), we
rewrite it as
\begin{equation}
  \label{eq:new-split-rewrite}
  \hG D(x,Q^2) = \int_0^1 \frac{dz}{z} P^{(1)}(z) 
  \left[e^{ \ln z \, \beta (\al) \partial_\al} \al \right]   \,
  \left[e^{\sigma  \ln z \,  \partial_{t}}
    D\left(\frac{x}{z},Q^2\right) \right] ,
\end{equation}
where both $\al$ and $D$ are now evaluated at scale $Q^2$ and
$\beta(\al) \equiv - d \al/d t = -\sum_{n=0} \beta_n \al^{n+2}$ .  The
Mellin transform of this equation results in the formal expression,
\begin{equation}
  \label{eq:new-split-mellin}
  \hG D_N = \gone\!\left(N + \beta (\al) \partial_\al + \sigma
  \partial_{t}\right) \,
   \al D_N  \,,
\end{equation}
where $\gone(N)$ is the Mellin transform of the first order splitting
function $P^{(1)}(x)$. We note that $\partial_{t}$ operates only on
$D_N$ and not on $\al$. This gives an all-order model for the
anomalous dimension, $\gamma(N) \equiv D^{-1}_N \hG D_N$.

Expanding (\ref{eq:new-split-mellin}) results in
\begin{eqnarray}
  \label{eq:new-split-mellin-expand}
\gamma\equiv   \gamma[\al] &=&    \al \gone + {\dgone} D^{-1} (\beta\partial_\al + \sigma \partial_{t}) (\alpha D) +
   + \half {\ddgone}\, D^{-1} (\beta\partial_\al + \sigma \partial_{t})^2 (\alpha D) + \ldots \nonumber \\
 &=&    \al \gone +  {\dgone} (\beta + \sigma \al \gamma)  
   + \half {\ddgone}\, D^{-1} (\beta\partial_\al + \sigma \partial_{t})
   (\beta D+ \sigma \al D \gamma)  + \ldots \nonumber \\
&=&    \al \gone +  {\dgone} (\beta + \sigma \al \gamma) + \half {\ddgone} \left[ \al \gamma^2 
+ \sigma(2\beta\gamma+\al\beta \partial_\al\gamma) + \beta\partial_\al \beta \right] + \cO{\al^4},
\end{eqnarray}
where dots indicate derivatives with respect to $N$ and $\gamma_1 \equiv
\gamma_1(N)$.  Solving this iteratively
produces
\begin{multline}
  \label{eq:iterative}
   \gamma 
   = \al \gone  + \al^2\,  \dgone (\beta_0  + \sigma\gone ) + \al^3\left[  \dgone(\beta_1+ \sigma\dgone(\beta_0+\sigma \gone ))
+ \half \ddgone  ( \gone^2 + 3\sigma\beta_0\gone + 2\beta_0^2)
   \right]+ \cO{\al^4}.
\end{multline}
For the purpose of understanding the MVV relation it suffices to take
$\gamma_1 = -A_1 \ln N + \cO{1}$ and to keep in (\ref{eq:iterative})
only the term $\propto \dgone\gone$, giving
\begin{equation}
  \label{eq:weJustGotC}
   \gamma = -\al A_1  \ln N + \mathrm{const.} + \sigma \al^2 A_1^2 \, \frac{\ln
     N}{N} + \cO{N^{-1}}.
\end{equation}
Recalling that $A_1^{a} \al  \equiv A^{a}$
(with $A_1^q = 4C_F$ and $A_1^g = 4C_A$), we can then write the
following all-order relation between $C^a$ and $A^a$,
\begin{subequations}
\begin{equation}
  \label{eq:C-all-orders}
  C^a = -\sigma (A^a)^2\,,
\end{equation}
or equivalently, in terms of the expansion coefficients $C_n$,
\begin{equation}
  \label{eq:Cn}
  C_1 = 0\,,\qquad
  C_2 = -  \sigma A_1^2\,,\qquad 
  C_3 = -2 \sigma A_1 A_2\,,\qquad 
  C_4 = -  \sigma (A_2^2 + 2A_1 A_3) \,,
  \qquad\mathrm{etc.}
\end{equation}
\end{subequations}
where we have suppressed the index $a=q,g$.  For the space-like case
($\sigma=-1$) this explains the MVV observation. For the time-like
case we have only the two-loop result \cite{CFP} to compare to, and it
agrees.  

\section{Pushing our luck}

Motivated by the idea that the universality of soft gluon emission holds
both in singular and constant terms in gluon energy \cite{LBK}, one
may attempt to trace further terms of the large-$N$ expansion
generated by eq.~(\ref{eq:new-split}). We definitely expect this push
to fail at the level of $1/N^2$ (possibly modulo logarithms, see
below) because this corresponds to `quantum' terms in the splitting
function, which vanish as $1\!-\! x$. However we would expect to have
control over the $1/N$ term in the anomalous dimension, 
\begin{equation}
\label{largeNN}
\gamma (N) = -A(\psi (N\!+\!1) +\gamma_e) 
            + B - C (\psi
            (N)+\gamma_e)\, N^{-1}  + D\, N^{-1} + \cO{N^{-2}\log^p N}\,.
\end{equation}
Compared to (\ref{MVVlargeN}) we have shifted the argument of the
logarithm in the $A$-term, $N\to N+1$, added the constant $\gamma_e$
in the $C$-term and then replaced logarithms with $\psi$ functions.
These modifications do not affect the first three functions $A$, $B$
and $C$ but serve to simplify the next subleading term $\propto 1/N$.
Additionally they lead to a compact $x$-space image of
(\ref{largeNN}),
\begin{equation}
\label{largeNx}
 P(x) =  \frac{A\, x}{(1\!-\!x)_+} + B\, \delta(1\!-\!x)\> +
 C\ln(1\!-\!x) \> +  D + \cO{(1\!-\!x)\log^p(1-x)}\,.
\end{equation}
Here, the presence of $x/(1-x)$ in the first term (as opposed to
$1/(1-x)$) is a consequence of Low's theorem \cite{LBK}.  In the
calculation of the $C$ coefficients we could safely ignore the
$\cO{1}$ piece of $\gamma_1$. This is no longer possible when
calculating $D$ because, in our non-linear construction, this constant
(\emph{unity} in Mellin space) is multiplied by $\dgone \sim 1/N$ thus
contributing to $D$. The extension of (\ref{eq:iterative}) to account
for $B$ in all orders is obtained by generalising
\begin{equation}
  \label{eq:general-wtih-B}
  \al \gone \to  \al
  (\gone - B_1) + B = -A(\ln (N\!+\!1) +\gamma_e) + B + \cO{N^{-2}}\,.
\end{equation}
The reason why the structure of the Taylor expansion
(\ref{eq:iterative}) is unmodified modulo this simple substitution is
that $B$, being a constant, disappears everywhere but un-dotted factors
of $\gamma_i$.

This leads to the following all-order expectation for $D$,
\begin{equation}
  \label{eq:D-all-order}
  D^a = A^a\left( \frac{\partial_t A^a}{A^a} -\sigma B^a \right)\,,
\end{equation}
where we have rewritten the $A_1^a \beta$ term that comes from
eq.~(\ref{eq:new-split-mellin-expand}) as $-(\partial_t A^a)$
(recalling the definition of $\beta$ in terms of the physical
coupling). The $\MSbar$ expansion for $D^a$ is then
\begin{equation}
  \label{eq:the-Ds-general}
   D_1=0,\quad  D_2 = -A_1(\sigma B_1 + \beta_0)\,,\quad
   D_3= -A_1(\sigma B_2 +\beta_1) - A_2(\sigma B_1 +\mathbf{2}\cdot\beta_0)\,,\quad
   \mbox{etc.} 
\end{equation}
\paragraph{Hard luck.} Examining the full known results for the two
and three-loop splitting functions, we find agreement for $D_2$ (space
and time-like, and quark and gluon channels); however the result for
the space-like $D_3$ is as follows (for both quarks and gluons)
\begin{equation}
  \label{eq:the-Ds-true}
   D_3= A_1(B_2 -\beta_1) + A_2(B_1 - \mathbf{1}\cdot\beta_0)\,.
\end{equation}
There is one mismatch between eqs.~(\ref{eq:the-Ds-general}) and
(\ref{eq:the-Ds-true}), which we have highlighted in boldface. Had
(\ref{eq:D-all-order}) contained $(\partial_t \almsbar)/\almsbar$
instead of $(\partial_t A^a)/A^a$ we would have obtained agreement
with the full result for $D_3$, however we see no reason why it should
be the $\MSbar$ coupling that appears there instead of the physical
coupling (which is equivalent to putting $A^a$). The remarkable
simplicity of the mismatch calls for a further \emph{structural
  explanation}.

\paragraph{Good luck.} Despite the disagreement in the comparison with
the exactly calculated subleading $D$ term (which we hope can be
understood) we have also investigated the coefficient of terms that
vanish for $x\to1$ but that are logarithmically enhanced there. We
have found agreement using (\ref{eq:iterative}) for the $\as^2 (1-x)
\ln(1-x)$ and the $\as^3(1-x)\ln^2(1-x)$ terms in the (space-like)
non-singlet anomalous dimensions, while $\as^3(1-x)\ln(1-x)$ contains
structures that remain to be understood. As for the diagonal singlet
anomalous dimensions, at three loops the coefficient of
$\as^3(1-x)\ln^2(1-x)$ agrees only in its $n_f$-independent parts and
there is an additional unexpected $\as^3(1-x)\ln^3(1-x)$ contribution
proportional to $n_f$, whose origin may only be explained once the
higher-order structure of the off-diagonal splittings is elucidated.

%
\medskip

The wealth of structure that is present in higher-order splitting
functions is suggestive of underlying simplicity. 
Possible sources of such simplicity, as proposed here, are the
universal nature of soft gluon radiation and the reformulation of the
notion of parton splitting functions with the aim of preserving
universality between space and time-like parton multiplication
(Gribov-Lipatov reciprocity).
Whether this picture can be made fully consistent remains to be seen.
We look forward to future work shedding more light on this question.

\section*{Acknowledgements}

We wish to thank Andreas Vogt for numerous discussions on the subject
of splitting functions as well as comments on the manuscript. The
comparisons performed here would not have been possible without the
explicit code provided in the arXiv versions of \cite{MVV,VMV} and
fortran \cite{FortranPolyLog} and mathematica \cite{MathPolyLog}
packages for dealing with harmonic polylogarithms.
We additionally thank Einan Gardi and Alberto Guffanti for a helpful
comment, and Werner Vogelsang for bringing to our attention
ref.~\cite{Stratmann:1996hn}. Its detailed study of Gribov-Lipatov
reciprocity in the second loop should help elucidate the structure of
non-diagonal transitions in eq.~(\ref{eq:new-split}).

\appendix
\section*{Appendix}

We do not know how to generalise~(\ref{eq:new-split}) to non-diagonal
transitions. These
are considerably more divergent at large $x$ \cite{FP} than would be
expected based on the non-linear relations that we propose here, going
as $\as^n \ln^{2n-2} (1-x)$ at large $x$.  We suspect that the origin
of these additional logarithms may be that in the $\MSbar$
factorisation scheme, for $a \to b$ transitions with $a \neq b$, the
splitting functions could pick up residues from ratios of
non-cancelling divergent Sudakov exponents (as well as from singular
integrals of these ratios). This belief is not inconsistent with the
MVV observation that in the supersymmetric case most of these
logarithmic enhancements cancel, since in this case the Sudakov
exponents become identical for quarks and gluons and do cancel.

These problems of non-diagonal terms may be responsible for the
following fact: at two loops, the $\cP$ that appears in
(\ref{eq:new-split}) for \emph{gluon-gluon} splitting is universal
(identical for space and time-like cases) only for two of the colour
structures, $C_A^2$ and $C_A n_f$. The analogue of (\ref{GLviol}) for
the remaining colour structure, $C_F n_f$ relates gluon-gluon and
singlet quark-quark splittings. On the left-hand side one finds the
combinations $P_{gg}^{(2),T} - P_{qq}^{(2),S}$ and $P_{qq}^{(2),T} -
P_{gg}^{(2),S}$, while on the right-hand side one has convolutions
involving $P_{gq}^{(1)}(x/z) \cdot \ln z P^{(1)}_{qg}(z)$ and
$P^{(1)}_{qg}(x/z) \cdot \ln z P^{(1)}_{gq}(z)$~\cite{Dunpub}.



\begin{thebibliography}{99}
\bibitem{DMSup} Yu.L.\ Dokshitzer, G.P.\ Salam and G.\ Marchesini, under preparation.
\bibitem{MVV} S.\ Moch, J.A.M.\ Vermaseren and A.\ Vogt, 
\np{688}{101}{2004}  
\bibitem{VMV}
A.\ Vogt, S.\ Moch and J.A.M.\ Vermaseren,  \np{691}{129}{2004} 
%
\bibitem{CFP} G.\ Curci, W.\ Furmanski and R.\ Petronzio,  \np{175}{27}{1980} 
%
\bibitem{KKST}
J.\ Kalinowski, K.\ Konishi, P.N.\ Scharbach and T.R.\ Taylor, 
\np{181}{253}{1981}
%
\bibitem{GL} V.N.\ Gribov and L.N.\ Lipatov, 
{Sov.\ J.\ Nucl.\ Phys.}~{\bf 15} (1972) 438,  
{\em ibid.}\/ {\bf 15} (1972) 675 

  
\bibitem{Dunpub} Yu.L.\ Dokshitzer, talk at the `Future Physics at
  HERA' Workshop, DESY Hamburg, 1995--1996, unpublished.


\bibitem{DKTheavy}
Yu.L.\ Dokshitzer, V.A.\ Khoze and S.I.\ Troian,
Phys.\ Rev.\ D {\bf 53}, 89 (1996)

 

\bibitem{LBK} F.E.\ Low,  
{Phys.\ Rev.}~{\bf 110} (1958) 974 ; \\ 
T.H.\ Burnett and N.M.\ 
Kroll,  
{Phys.\ Rev.\ Lett.}~{\bf 20} (1968) 86
%
\bibitem{DS} Yu.L.\ Dokshitzer and D.V.\ Shirkov, 
{Z.\ Phys.}\/ {\bf C67} (1995) 449 


\bibitem{FortranPolyLog}
  T.~Gehrmann and E.~Remiddi,
  Comput.\ Phys.\ Commun.\  {\bf 144} (2002) 200
  [hep-ph/0111255].

\bibitem{MathPolyLog}
  D.~Maitre,
  ``HPL, a Mathematica implementation of the harmonic polylogarithms,''
  hep-ph/0507152.

\bibitem{FP}
  W.~Furmanski and R.~Petronzio,
  Phys.\ Lett.\ B {\bf 97} (1980) 437.

\bibitem{Stratmann:1996hn}
  M.~Stratmann and W.~Vogelsang,
  Nucl.\ Phys.\ B {\bf 496} (1997) 41
  [arXiv:hep-ph/9612250].

\end{thebibliography}
\end{document}